\documentclass[twocolumn,showpacs,preprintnumbers,amsmath,amssymb]{revtex4}

\usepackage{graphicx}
\usepackage{dcolumn}
\usepackage{bm}
\usepackage{subfigure}

\begin{document}

\title{Quantum memories with electrically controlled storage and retrieval in an opto- and electro-mechanical cavity}

\author{Li-Guo Qin$^1$}
\email{qinlg@sari.ac.cn}
\author{Zhong-Yang Wang$^{1}$}
\email{wangzy@sari.ac.cn}
\author{Gong-Wei Lin$^2$}
\author{Jing-Yun Zhao$^1$}
\author{Shang-Qing Gong$^{2}$}
\email{sqgong@ecust.edu.cn}
\affiliation {$^1$Shanghai Advanced Research Institute, Chinese Academy of Sciences, Shanghai, 201210, China \nonumber\\
$^2$Department of Physics, East China University of Science and Technology, Shanghai 200237, China}

\date{\today}

\begin{abstract}

We propose a novel scheme to realize electrically controlled quantum memories in the opto- and electro-mechanical (OEM) cavity. Combining this OEM cavity with the mechanism of Electromagnetically Induced Transparency (EIT) we find that the quantum interference, arising from the two optical transitions of the $\Lambda$ type three-level atomic ensembles, can be manipulated electrically. Numerical calculations show that the probe photon state can be well stored into the atomic spin state by sending an electric current pulse and retrieved with time-reverse symmetry by sending the other current pulse with opposite direction. The quantum interference with electric controlling is expected to apply to other quantum control aspects.


\end{abstract}

\maketitle

The storage and retrieval of quantum state, i. e., quantum memory(QM), is the central core of the quantum technology \cite{pws,mfm,ail,gnm,bjc}. Successful QM schemes, such as electromagnetically induced transparency (EIT) \cite{mda}, Raman \cite{aek} and photon-echo \cite{sam}, use the optically controlled quantum interference effects to transfer the photon state (flying qubit) to spin state. However, the fragility of quantum interference makes the progress of practical application of QM slow and difficult \cite{lmd,nsc}. On the other hand, quantum electromechanical systems \cite{mbl}, which consist of nano-to-micrometer scale mechanical resonators coupled to electronic devices of comparable dimensions, have been widely employed in quantum thermal conductance of suspended dielectric wires \cite{kse}, displacement transducer of the single electron transistor \cite{rka} and coherent control device of the cooper-pair box \cite{dva}, etc. Using such systems, the detection sensitivity to force, mass and displacement approaches the quantum limit \cite{mbl,kcs}. Recently, electrically controlled methods, such as electrically driven single photon emission \cite{yuan} and entangled photon pair generation \cite{cls} and source \cite{mbw,thc}, and electrically pumped polariton light-emitting device \cite{sit}, have been experimentally demonstrated. These electrically controlled quantum devices present the significant breakthrough in quantum applications. In this letter, we propose a novel scheme to achieve the electrically controlled quantum interference effect in the optomechanical cavity, therefore the photon state can be stored and retrieved through the electrical control.

The key of our scheme is to use optomechanical cavity with electric charge controlled mechanical oscillator. In recent years, the optomechanical system has made big progress as well as the electromechanical system. Through the coherent coupling between the optical and mechanical degree of freedom by the radiation pressure inside the cavity, the mechanical oscillator can be cooled down to quantum ground state \cite{ado,jdt,jct,nbt}. The optomechanically induced transparency (OMIT) can be realized by tuning the control field resonant with one of the side band transition of optomechnical system \cite{swr}, in which EIT, normal mode splittings, quantum state conversion and pulse transmission have also been observed \cite{swr,ahs,gsa,ltn}. Based on OMIT, a scheme of precision measurement of electrical charge has been proposed \cite{jqz}, in which the charged mechanical oscillator coupled to a nearby charged object by the coulomb force has been first used. Taking advantage of cavity and the charged mechanical oscillator, in this letter, we design an opto- and electro- mechanical (OEM) cavity with the optomechanically and electrically controlled displacement of the moveable mirror (MR). By combining cavity optomechanics and EIT mechanism, QM with electrically controlled photon storage and retrieval is thus proposed.

In our scheme the spin wave of the atomic ensemble is used as the storage medium owing to the long decoherence time, which is trapped inside a high-Q optical cavity. With the destructive quantum interference induced by the cavity control field, an opaque resonant atomic medium becomes transparent for a probe field. In the high-Q cavity the coupling between atomic medium and the cavity field can be greatly enhanced. When the interaction of matter and cavity field enter the strong coupling regime, i.e., the coherent coupling rate exceeds the decoherence rate of each of the subsystems \cite{sgk}, the transparency can be induced by few photons and even by cavity vacuum. Such cavity and vacuum induced transparency have been explored in recent years, and can be used as photon number-state filters and photon number -resolving detectors \cite{gnm,prr,hts}.

In our model the MR of OEM cavity acts as the mechanical oscillator with the frequency $\omega_m$ and the mass $m$, which is driven by the radiation pressure induced by cavity field with resonance frequency $\omega_0$. This mechanical oscillator is charged by electron voltage and coupled to the other opposite charged object by coulomb force, therefore the mechanical oscillator can be electrically controlled. As shown in Fig. 1, we consider a uniform atomic medium of length $l$ and cross-section area $S$ trapped inside an OEM cavity. Here the quantized cavity field $\hat{a}$ couples to atomic dipole transition between level $|c\rangle$ and $|a\rangle$ with a coupling strength $g$ and also to the mechanical oscillator through the radiation pressure. The photon is resonantly injected into the cavity by an optical field $E_c$ from the fixed cavity mirror in order to sustain the average photon number inside the cavity. The stored photon state or a weak probe field with frequency $\omega_p$ externally injects into the cavity along the $z$ axis, and couples to the atomic dipole transition between level $|b\rangle$ and $|a\rangle$ with a coupling strength $g_p$.

Hence the total Hamiltonian of the system including the interactions among the externally optical field $E_c$ and photon state $\hat{E}_p$, the cavity field $\hat{a}$, the atoms, the mechanical oscillator and the charged object is given by

\begin{figure}[h]
\includegraphics[angle=0,width=8cm]{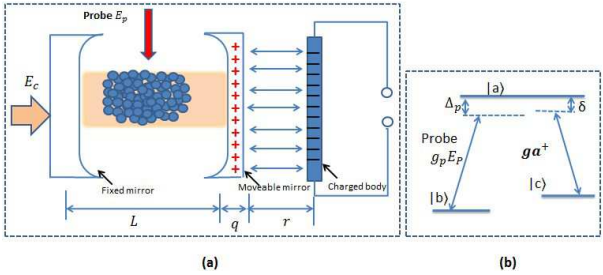}
\caption{(Color online) Schematic diagram of the system: (a) The OEM setup, standard optomechnical cavity with the electromechanical mirror, is driven by the injection optical field $E_c$ and the electric field. An atomic ensemble of $N$ identical $\Lambda$ type three-level atoms confined in the cavity with length $L$ coupled to a probe field $E_p$ and the cavity field $g\hat{a}$. (b) The level structure of the $i$th atom coupling to two optical fields. The decay rate $\gamma_1$ and $\gamma_3 (\gamma_2)$ from $|a\rangle$ to $|b\rangle$ and$|c\rangle$ (from $|c\rangle$ to$|b\rangle$) are phenomenologically introduced, $\Delta_p=\omega_{ab}-\omega_p$ and $\delta=\omega_{ac}-\omega_0$ in the coupling scheme.}
\end{figure}

\begin{eqnarray}
H=H_{0}+H_{I},
\end{eqnarray}
where $H_0$, the free hamiltonian of the model, is given explicitly by
\begin{eqnarray}
H_{0}=&&\hbar\omega_{0}\hat{a}^{\dagger}\hat{a}+(\frac{\hat{p}^2}{2m}+\frac{1}{2}m\omega_{m}^2\hat{q}^2)\nonumber\\&&+\hbar\omega_{ab}\sum_{i=1}^{N}\hat{\sigma}^{(i)}_{aa}+\hbar\omega_{cb}\sum_{i=1}^{N}\hat{\sigma}^{(i)}_{cc},
\end{eqnarray}
which represents the steady hamiltonians of the energy of the cavity mode, MR and atomic ensemble. $\hat{a}$ ($\hat{a}^\dag$) is the annihilation (creation) operator for the cavity mode, $\hat{q}$ and $\hat{p}$ are the position and momentum operators of the MR, respectively; $\hat{\sigma}^{i}_{\alpha\beta}=|\alpha\rangle_{ii}\langle\beta|$ ($\alpha,\beta=a,b,c$) are the atomic operators; here, $\hbar\omega_{ab}(\omega_{cb})$ is the energy-level spacing between $|a\rangle_i$ and $|b\rangle_i$ ($|c\rangle_i$ and $|b\rangle_i$) for the $i$th atom. $H_I$, the interaction hamiltonian, accounts for the coupling of the various subsystems
\begin{eqnarray}
H_{I}=&&-\hbar G_{0}\hat{a}^{\dagger}\hat{a}\hat{q}+i\hbar\varepsilon_{c}(\hat{a}^{\dag}e^{-i\omega_{c}t}-\hat{a}e^{i\omega_{c}t})
-\sum_{i=1}^{N}\hbar (g_{p}\hat{E}_{p}\nonumber\\&&\hat{\sigma}^{(i)}_{ab}+g\hat{a}\hat{\sigma}^{(i)}_{ac}+H.c.)-\frac{n_q|e|Q_{mr}}{4\pi\varepsilon_0(r-q)}.
\end{eqnarray}
The first term gives the coupling of the cavity field to the mechanical oscillator of MR with the optomechanical coupling rate $G_0=\frac{\omega_c}{L}$. The second term corresponds the input-output relations of the cavity field and the injection optical field $E_c$ with frequency $\omega_c$ and strength $\varepsilon_c$ related to the laser power $P_c$ by $\varepsilon_c=\sqrt{2P_{c}\kappa/\hbar\omega_c}$, where $\kappa$ is the cavity decay rate. The third term describes the two dipole interaction between atomic transitions $\sigma^{i}_{ab}$ ($\sigma^{i}_{ac}$) and the probe field $\hat{E}_p$ (the cavity field $\hat{a}$) with the coupling strength $g_p=\wp_{ab}\sqrt{\omega_p/(2\hbar\varepsilon_0 V)}$ ($g=\wp_{ac}\sqrt{\omega_c/(2\hbar\varepsilon_0 V)}$), $\wp_{ab}$ ($\wp_{ac}$) is the dipole matrix element of transitions between $|a\rangle$ and $|b\rangle$ ($|c\rangle$) and $V$ is the effective interactive volume of the atom-photon. The last term represents the interaction of the MR with the charged body via a Coulomb potential $V_c$, where $Q_{mr}$ is the positive charge on the MR and $-n_q|e|$ is the negative charge of the charged body with charge number $n_q$. Here the coulomb force on the MR points to the same direction as the radiation pressure force on the MR. Because of $q\ll r$, the Coulomb interaction can be approximated as $V_c\approx-\frac{n_q|e|Q_{mr}}{4\pi\varepsilon_0 r}(1+\frac{q}{r})$ \cite{jqz}. Let $\eta=\frac{|e|Q_{mr}}{4\pi\varepsilon_0 r^2}$, the last term can be rewritten as $-n_q \eta (q+r)$.

To describe the propagation of the probe pulse through the atomic ensemble, we use Maxwell-Bloch equations in the slowly
varying amplitude approximation. Thus the propagation of the probe pulse can be obtained by \cite{gnm}
\begin{eqnarray}
(\frac{\partial}{\partial z}+ \frac{\partial}{c \partial t})\hat{\alpha}=i \frac{ g_p^2 N}{c} \hat{\sigma}_{ba}= i \eta_p \hat{\sigma}_{ba},
\end{eqnarray}
where $\hat{\alpha}=g_p \hat{E}_p$, $\hat{E}_p =\hat{E}(t) e^{-i\omega_{p}t}$ and $\eta_p=\frac{ g_p^2 N}{c}$. In addition, we use the collective operators of the atomic ensemble, $\sigma_{\alpha\beta}=\frac{1}{N}\sum_{i=1}^{N}\sigma^{(i)}_{\alpha\beta}$. Let $\gamma$, the decay rate of the upper energy level $|a\rangle$, as the characteristic time of the pulse, we redefine the dimensionless: $\hat{a} /\gamma\rightarrow \tilde{a}$, $\hat{\alpha}/\gamma\rightarrow \tilde{\alpha}$, $\delta /\gamma\rightarrow\tilde{ \delta}$, the dimensionless time $t\gamma\rightarrow t^{'}$, and the length $z\eta_p\gamma\rightarrow z^{'}$. In the following a comoving frame $\tau=t^{'}-z^{'}/c$ and $\xi=z^{'}$ \cite{avg}, is used and obtained as $\partial_{\tau}=\partial_{t}$ and $\partial_{\xi}=\partial_{z}+\frac{1}{c}\partial_{t}$. Thus the dynamics of the system governed by the Hamiltonian $H$ in Eq. (1) are obtained as
\begin{subequations}
\label{allequations}
\begin{eqnarray}
&&\partial_{t}\hat{q}=\frac{\hat{p}}{m},\label{equationa} \\
&&\partial_{t}\hat{p}=-m\omega_{m}^{2}\hat{q}-\gamma_{m}\hat{p}+\hbar G_{0}\hat{a}^{\dag}\hat{a}+n_q\eta,\label{equationb}\\
&&\partial_{t}\hat{a}=[-\kappa-i(\Delta-G_{0}q)]\hat{a}+\varepsilon_{c}+igN\hat{\sigma}_{ca},\label{equationc} \\
&&\partial_{\tau}\hat{\sigma}_{aa}=-2\hat{\sigma}_{aa}+i (\tilde{\alpha}\hat{\sigma}_{ab}-H.c.)+i (\tilde{\Omega}\sigma_{ac}-H.c.),\;\;\;\;\;\;\;\label{equationd}\\
&&\partial_{\tau}\hat{\sigma}_{bb}=\hat{\sigma}_{aa}-i (\tilde{\alpha}\hat{\sigma}_{ab}-H.c.),\label{equatione}\\
&&\partial_{\tau}\hat{\sigma}_{cc}=\hat{\sigma}_{aa}-i (\tilde{\Omega}\hat{\sigma}_{ac}-H.c.),\label{equationf}\\
&&\partial_{\tau}\hat{\sigma}_{ab}=-(1-i\tilde{\Delta}{p})\hat{\sigma}_{ab}+i \tilde{\alpha}^{\dag}(\hat{\sigma}_{aa}-\hat{\sigma}_{bb})-i\tilde{\Omega}^{\dag}\hat{\sigma}_{cb},\label{equationg}\\
&&\partial_{\tau}\hat{\sigma}_{cb}=i(\tilde{\Delta}{p}-\tilde{\delta})\hat{\sigma}_{cb}+i \tilde{\alpha}^{\dag}\hat{\sigma}_{ca}-i\tilde{\Omega}\hat{\sigma}_{ab},\;\;\;\;\;\;\;\label{equationh}\\
&&\partial_{\tau}\hat{\sigma}_{ca}=-(1+i\tilde{\delta})\hat{\sigma}_{ca}+i \tilde{\alpha} \hat{\sigma}_{cb}-i\tilde{ \Omega}(\hat{\sigma}_{aa}-\hat{\sigma}_{cc}),\label{equationi}
\end{eqnarray}
\end{subequations}
where $\gamma_m$ is the damping of MR, $\Delta=\omega_0-\omega_c$ and $\hat{\Omega}=g \hat{a}$ and the notation $\tilde{X}=\hat{X}/\gamma$. Here we assume the decay $\gamma_{ab}=\gamma_{ac}=\gamma$ and consider the case that most occupations of atoms are initially prepared in the ground state, i.e., $\sigma_{bb}(t=0)\approx 1$, and the population in the other states are almost zero. In our model, the Coulomb force acting on MR is much bigger than the radiation pressure, and the damping rates of MR and cavity ($\gamma_m$ and $\kappa$) are much longer than the characteristic time $\Gamma=1/\gamma$. When the charge is slowly injected into the charge object, the evolution of $\hat{q}$, $\hat{p}$ and $\hat{a}$ can be considered adiabatically, namely the time derivatives of Eqs. (5a-5c) can be neglected. In other words, under the adiabatic limit the MR and cavity are regards as in the steady states during the interaction of the photons and atoms. In addition, $\sigma_{ca}$ in Eq. (5c) can also approximate as zero due to the much larger density of atoms than the probe photons. With these conditions and Eqs. (5a-5c), We can obtain the equations for the MR position $\hat{q}$,
\begin{subequations}
\begin{eqnarray}
&& m\omega_{m}^2\hat{q}-\hbar G_{0}\hat{n}=n_q \eta,\label{equationa}\\
&& \hat{n}=\hat{a}^{\dag} \hat{a}=\frac{|\varepsilon_{c}|^2}{\kappa^2+(\Delta-G_{0}\hat{q})^2},\label{equationb}
\end{eqnarray}
\end{subequations}
where $\hat{n}$ is the operator of the photon number in the OEM cavity. From the equations we can see that the charge $n_q$ largely affects the operators of photon number $\hat{n}$ and MR position $\hat{q}$, and the change of MR position alters the photon number inside the cavity. Thus the charge can manipulate the coherent coupling between the atoms and the cavity field. By changing the charge number the interaction of atoms and cavity can be dynamically controlled. Substituting Eq. (6b) to Eq. (6a), we can obtain the third-order nonlinear equation of the MR position operator, which has the analytical solutions about the relations among the MR position, the photon number and the charge number. To get a clear picture of the role of charge, we simplified the solution of MR position and charge number as $\hat{q}\approx\frac{n_q \eta}{m\omega_{m}^2}$, when $\hbar G_{0}\hat{n} \ll n_q \eta$, which is the case in our model, and $\hat{n}=\frac{|\varepsilon_{c}|^2}{\kappa^2+(G_{0}\hat{q})^2}$ when $\Delta=0$. So the MR position approximately change linearly with the charge number, but the photon number is inversely proportional to the square of MR position, that means the photon number inside the cavity is more sensitive with the change of MR position or charge number. With these relations in mind, in the following we show how these can be used as the dynamical control of quantum interference in atomic $\Lambda$ level system in the cavity.

It is known that the EIT scheme of QM use the optically controlled quantum interference effect. By dynamically and adiabatically turning off the control optical field, the single photon state can be stored in the spin state of atomic ensemble through the dark state polariton. After the storage, the single photon state can be read out on demand by adiabatically turning on again the control optical field. Here in our model the cavity field is used as the control field and controlled electrically. Fig. 2(a) shows how this dynamical process works. When the charge number in charge object increase, the deviation of the MR position follows the charge trend linearly, but the photon number of the cavity field decreases rapidly down to zero. Thus we can dynamically turn off the cavity field to realize the storage of photon state. Inversely, when the charges decrease, the deviation of the MR position decreases, and the photon number inside the cavity increases. The cavity field can turn on again, hence the photon state can be read out.

\begin{figure}
\subfigure[]{ \label{fig:a}
\includegraphics[width=0.8\columnwidth]{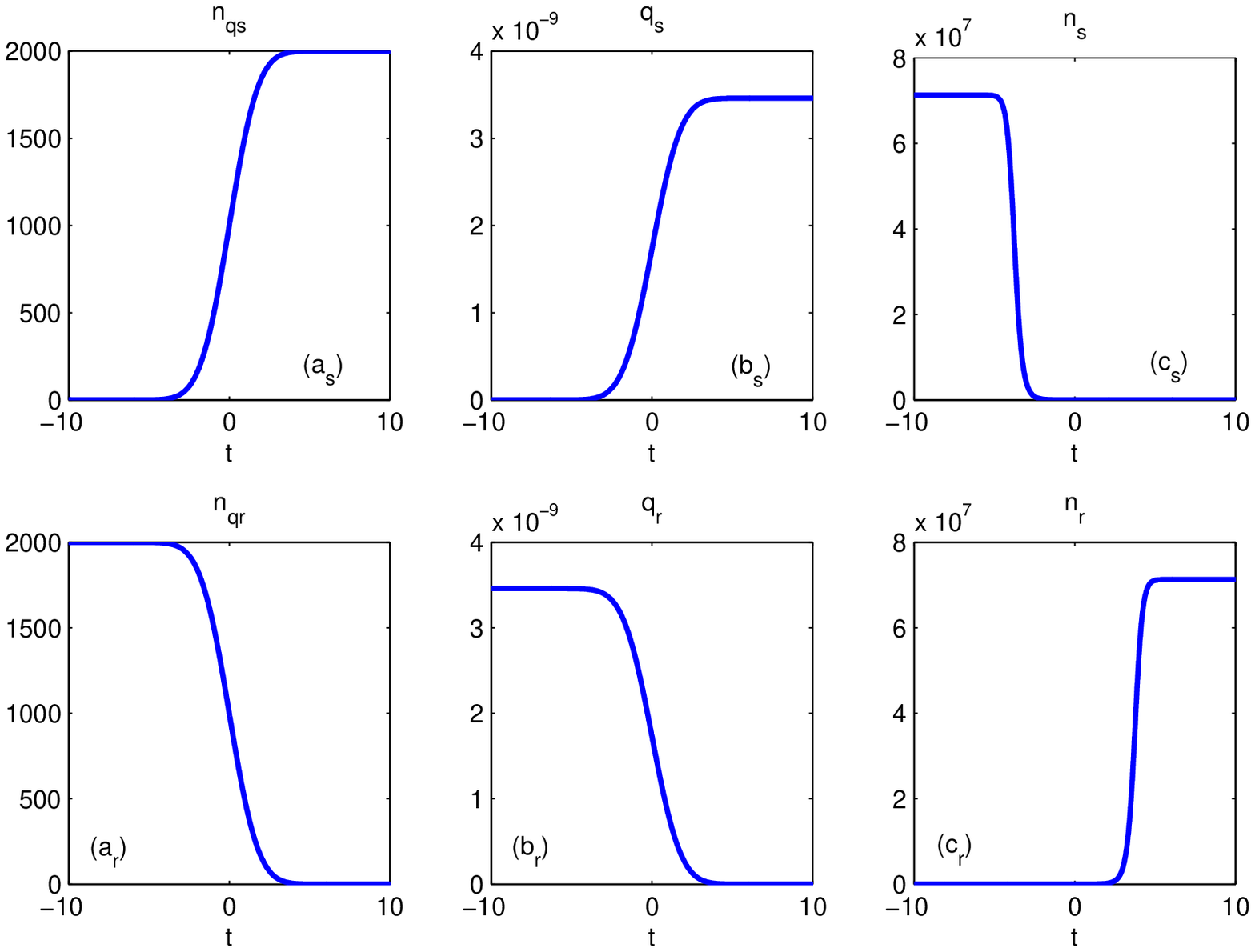}
}
\subfigure[] { \label{fig:b}
\includegraphics[width=0.5\columnwidth]{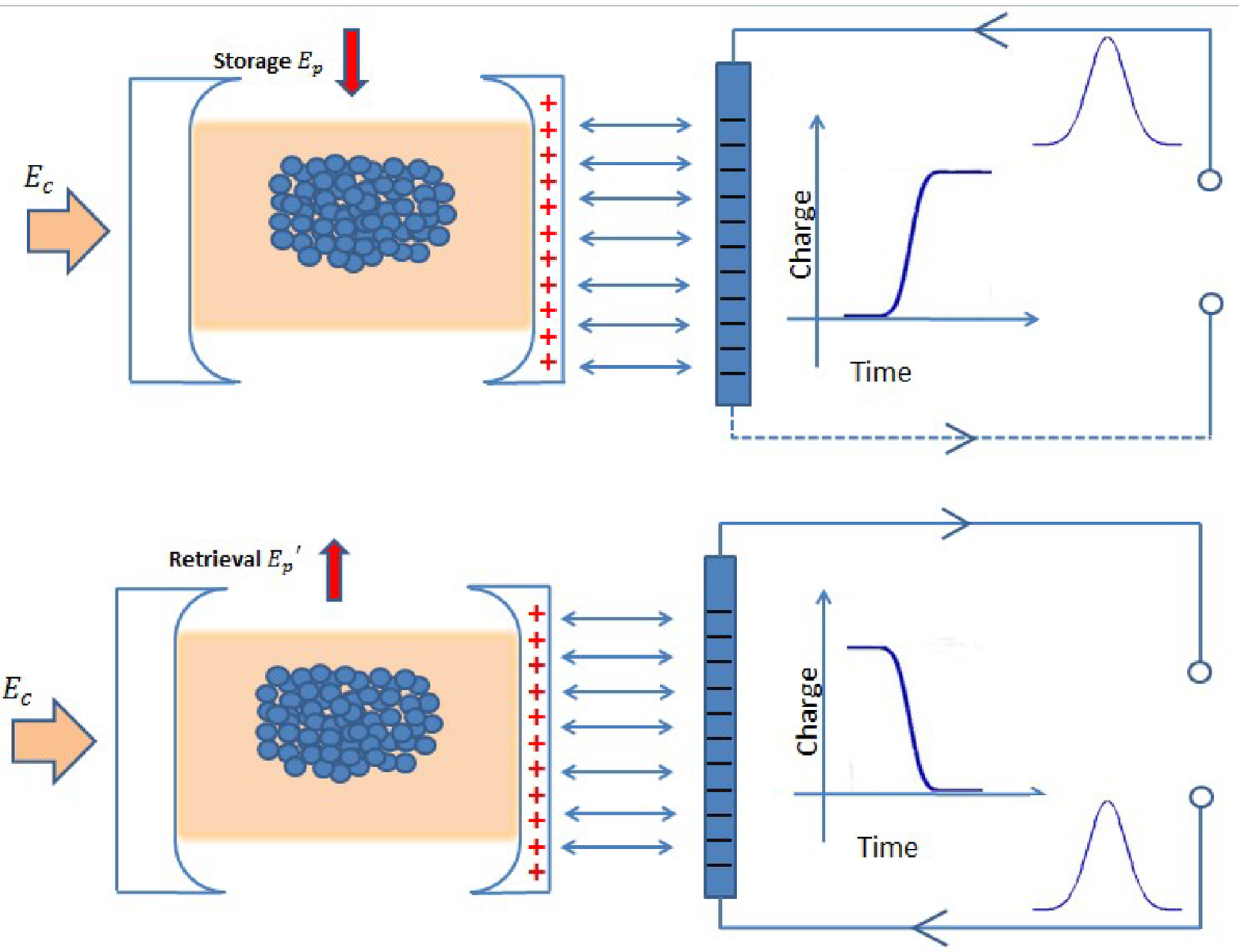}
}
\subfigure[] { \label{fig:c}
\includegraphics[width=0.4\columnwidth]{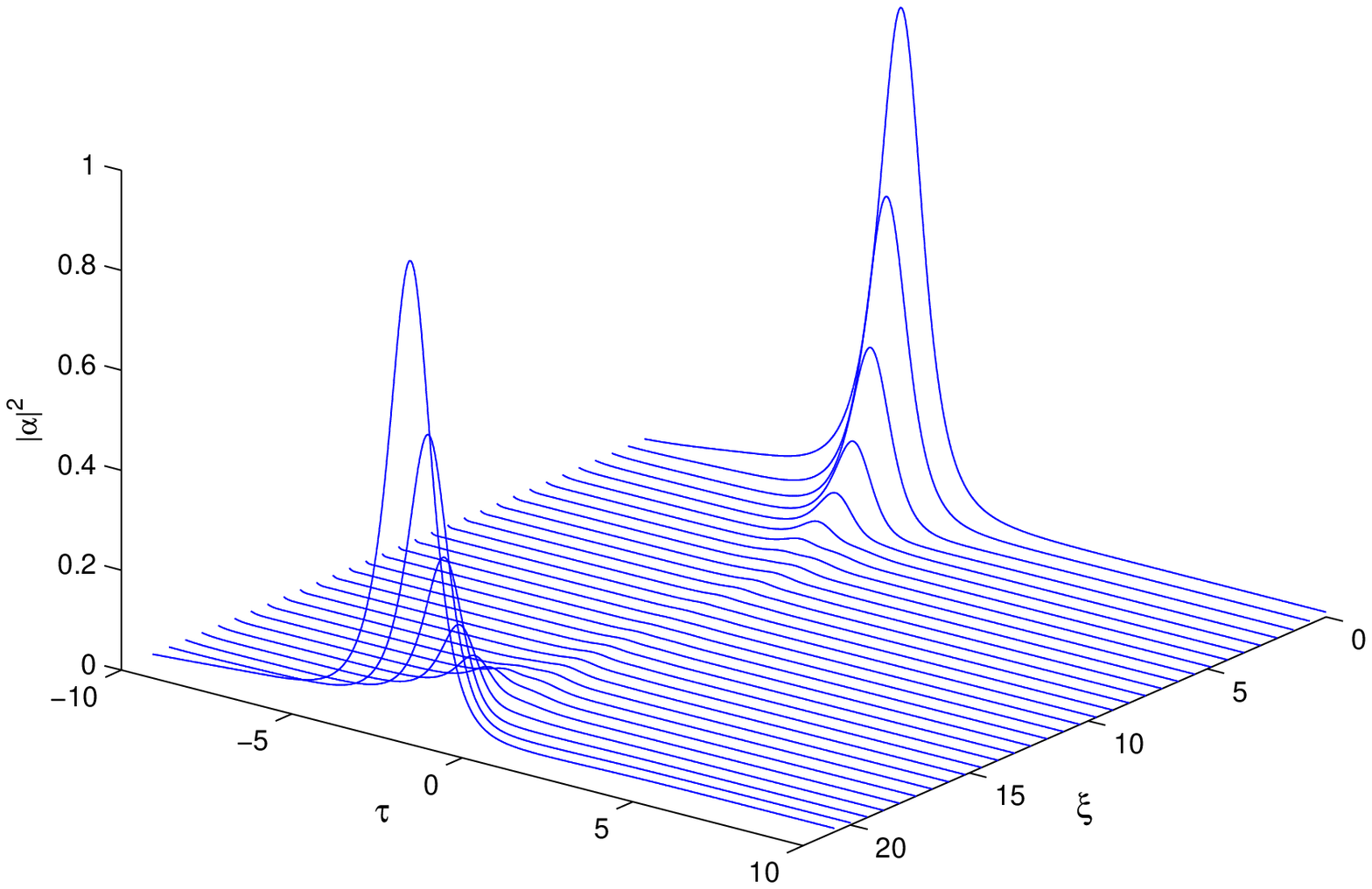}
}
\caption{The schematic of the storage and retrieval process with electric controlling. (a)$a_s$, $b_s$ and $c_s$ illustrate the evolution of charges, position of MR and the cavity photon number versus the time in the storage process, respectively; $a_r$, $b_r$ and $c_r$ show the corresponding evolutions versus the time in the retrieval process. (b)The diagrammatic drawings of the storage and retrieval process. (c) The result of the numerical simulation: the propagation of the probe field with the electrically controlled storage and retrieval is optimized by using the method (J. Nunn et al \cite{jni}) in the case of $\Delta_p=\delta=0.1\gamma$. The parameters are taken from the experiments in \cite{jqz,sgk} as $\lambda_c=795nm$, $L=25mm$, $m=145ng$, $\omega_m=2\pi\times947kHz$, $\gamma_m=2\pi\times141Hz$, $\kappa=2\pi\times215kHz$, $r=67\mu m$, $\varepsilon_c=\sqrt{2P_c\kappa/(\hbar\omega_c)}$ with $P_c=12\mu W$, $V=8mm^3$, $Q_{mr}=CU$ with $C=27.5nF$ and $U=1V$.}
\label{fig}
\end{figure}

To reveal the dynamical evolutions of storage and retrieval, we numerically solve Eq. (4) and Eqs. (5d-5i). As shown in the schematic figure of our QM scheme of Fig.2(b), an optical field is firstly injected into the cavity resonantly, the mechanical oscillator of MR is then excited by the radiation pressure of photons inside the cavity. Before the arriving of the stored photon, the cavity field also acts as the pumping field so that the population of atomic ensemble is optically pumped into the ground state $|b\rangle$. Afterwards, a wavepacket of the stored photon or the weak probe pulse $\hat{E}_{p}$ enters the cavity along the $z$ axis and interacts with the atomic medium. Simultaneously, the object is charged anticlockwise by a synchronized electric current pulse (here we take the Gaussian distribution of the pulse) in order to adiabatically turn on the charge as shown in Fig. 2(a)($a_s$), so that the photon number inside the cavity has been manipulated down to zero during the interaction of the stored photon and atomic medium, and maps the photon state to the atomic spin state through adiabatically transferring the dark-state polaritons in EIT scheme. After the storage, the photon state can be read out by time-reverse symmetry. Here this is achieved by charging the object with the Gaussian electric current pulse along the clockwise direction, which neutralizes the charge object down to zero and turns dynamically on the cavity field as shown in Fig. 2(a)($a_r$, $b_r$ and $c_r$). In this way, the stored photon is read out and leaves the cavity along the opposite direction. Hence the photon storage and retrieval can be achieved by the electrically controlled quantum interference effect of the atomic ensemble inside the optomechanical cavity. A probe pulse is stored into the atomic medium by simultaneously sending an electric current pulse. Subsequentially, the probe photon pulse is read out by the time reverse of retrieval, which can be realized by sending the other current pulse with opposite direction. However it is found that the mode of the probe pulse does not well match with the mode of spin state in the medium as discussed by J. Nunn et al \cite{jni} and Gorshkov et al \cite{avg}. J. Nunn et al. presented a mode matching theory in which the phase matching problem is discussed in the limit of vanishing Stokes shift and the dynamics are optimized using a universal mode decomposition. Gorshkov et al. presented a detailed theory of the EIT-based storage of light for a variety of experimental configurations, which has provided techniques for optimizing memory performance \cite{nbp,ina,inn}. Based on their techniques, we can obtain the excellent storage and retrieval of the pulse in the medium by sending the optimal shape of the probe pulse as shown in Fig. 2(c).

To illustrate the storage and retrieval efficiency of the probe pulse, we use the analysis presented by Gorshkov et al. \cite{avg,avg1,avg2}. Assuming that almost all atoms are in the ground state at all times $\sigma_{bb}\approx 1$ and $\sigma_{aa}\approx \sigma_{cc} \approx \sigma_{ca} \approx 0$, we redefine the polarization operator $P=\frac{1}{\sqrt{N}}\sum_{i=1}^{N}\sigma^{(i)}_{ab}$, the spin-wave operator $S=\frac{1}{\sqrt{N}}\sum_{i=1}^{N}\sigma^{(i)}_{cb}$ and the input field operator $\varepsilon=\sqrt{c/(l\gamma)}E_p$. Thus Eqs. (4) and (5d-5i) can reduce to $\partial_{\xi}\varepsilon=i \sqrt{d}P$, $\partial_{\tau}P=-(1-i\tilde{\Delta}{p})P - i\sqrt{d}\varepsilon -i\tilde{\Omega}^{\dag}S$, and $\partial_{\tau}S=-i\tilde{\Omega} P$, where $d=g^2Nl/(c\gamma)$ is the optical depth. With the two photon resonant, the normalization, initial and boundary conditions, the storage efficiency $\eta_s = \int^{l}_{0}d\xi|S(\xi,T)|^2$ and the retrieval efficiency $\eta_r = \int^{\infty}_{T}d\tau|\varepsilon(l,\tau)|^2$ for a smooth storage pulse of duration $T$ can be obtained. These give the same results with the previous EIT scheme, the difference is just the control field in place of the electrically controlled cavity field. Hence in our scheme the optimal storage and retrieval efficiencies can be reached and still depend on the optical depth for the optimal spin wave \cite{avg2}.

In QM schemes of the far off-resonant Raman scheme and the resonant EIT scheme the adiabatic condition is needed (as $Td\gamma \gg 1$) to adiabatically eliminate the excited state for the probe and control fields with the duration $T$. In photon-echo scheme, the adiabatic condition are considered that the excitations are mapped from the unstable excited state into the stable ground state and achieved by applying a fast and short resonant $\pi$-pulse ($T\sim 1/(d\gamma)$) at the right time. In our scheme the memory process for the probe and control fields should satisfy the same adiabatic condition. However, in our scheme the cavity field as the control field  requires the conditions $|\frac{\dot{a}}{a}| \ll |\gamma d+i\Delta_p|$ and $|a| \ll |\gamma d+i\Delta_p|$ for the dynamic storage and retrieval. Because the cavity field can be controlled by the electric current pulse, that means this adiabatic conditions can also be adjusted electrically. This adjustment has two approaches. One is the adjustment of the charge slope in Fig. 2(a) by using sufficiently slow varying electric current pulse. Another one is the adjustment of the distance between the MR and the charge object in the coulomb force. Because the coulomb force is inversely proportional to the square of the distance, so $q$ and $n$ is more sensitive with the distance between the MR and the charge object. Both two methods can affect the slope of $q$, especially $n$ in Fig. 2(a). These approaches have the advantages and convenience for realizing the adiabatical conditions in practice.

For the realizability of the proposed scheme, we consider the optomechanical cavity given by S. Gr\"{o}blacher et al in experiment \cite{sgk}, the cold atoms $^{87}Rb$ with $|a\rangle=|5^2P_{1/2},F=2\rangle$, $|b\rangle=|5^2S_{1/2},F=1\rangle$ and $|c\rangle=|5^2S_{1/2},F=2\rangle$ \cite{mda} trapped inside such optomechanical cavity \cite{fbe}. For the electromechanical oscillator we can use a suitable scale mechanical resonator for our model \cite{mbl}. The adiabatic limits can be meet by choosing the suitable the distance between the MR and the charge object and the width of the Gaussian electric current pulse. Hence the real electrically controlled quantum memories is experimentally feasible.

In conclusion, combining this OEM cavity with EIT scheme, in this letter a QM scheme with electrically controlled storage and retrieval is proposed. We find that the photons inside the OEM cavity can be manipulated by an electric current pulse putting on a charge object, which couple to the optomechanical oscillator by coulomb force. When an atomic ensemble with three levels $\Lambda$ system is trapped inside this OEM cavity, the quantum interference effect is then controlled electrically. By numerical calculation, we find that the probe photon state can be well stored into the atomic spin state by sending an electric current pulse, after the storage, the photon state can be read out with time-reverse symmetry by sending the other current pulse with opposite direction. We find that this scheme can achieve the same efficiency of previous EIT scheme. Especially, in our scheme the adiabatic conditions can be adjusted electrically by changing the width of the current pulse or the distance between MR and the charge object. We believe our electrically controlled QM scheme has the significance for the development of quantum key device. We also expect this scheme to be applicable to other quantum interference effects.

We acknowledge Simon-Pierre Gorza with the help of the programming in the Universit\'{e} libre de Bruxelles. This work is supported by the Strategic Priority Research Program of the Chinese Academy of Sciences, Grant No. XDB01010200, the Hundred Talents Program of the Chinese Academy of Sciences, Grant No. Y321311401 and National Natural Sciences Foundation of China (Grants No. 11204080 and No. 11274112).

\end{document}